\documentclass[11pt]{article}
\newcommand{\nc}{\newcommand}
\nc{\bib}{\bibitem}
\nc{\al}{\alpha}
\nc{\g}{\gamma}
\nc{\G}{\Gamma}
\nc{\D}{\Delta}
\nc{\eps}{\epsilon}
\nc{\la}{\lambda}
\nc{\La}{\Lambda}
\nc{\var}{\varphi}
\nc{\hn}{h^\vee}
\nc{\kn}{k^\vee}
\nc{\adg}{a^\dagger}
\nc{\bdg}{b^\dagger}
\nc{\ba}{\beta_\al}
\nc{\ga}{\g^{\al_1}}
\nc{\vpp}{{V_+}^+}
\nc{\cpp}{{C_+}^+}
\nc{\Vm}{V_{-\al^-}^{\al_1}}
\nc{\Vp}{V_{-\al^+}^{\al_1}}
\nc{\Vmb}{V_{-\beta^-}^{\al_1}}
\nc{\Gb}{\overline{G}}
\nc{\Gbc}{\overline{{\cal G}}}
\nc{\pa}{\partial}
\nc{\nn}{\nonumber \\ }
\nc{\hf}{\frac{1}{2}}         
\nc{\dz}{\frac{dz}{2\pi i}}
\nc{\fabc}{{f_{a,b}}^c}
\nc{\bin}[2]{\left (\begin{array}{c} {#1}\\ {#2} \end{array}\right )}
\nc{\ben}{\begin{equation}}
\nc{\een}{\end{equation}}
\nc{\bea}{\begin{eqnarray}}
\nc{\eea}{\end{eqnarray}}
\nc{\bra}[1]{\langle {#1}|}
\nc{\ket}[1]{|{#1}\rangle}
\newcommand{\Z}{\mbox{$Z\hspace{-2mm}Z$}}
\nc{\C}{\mbox{\hspace{1.24mm}\rule{0.2mm}{2.5mm}\hspace{-2.7mm} C}}
\nc{\Nat}{\mbox{\hspace{.04mm}\rule{0.2mm}{2.8mm}\hspace{-1.5mm} N}}
\newcommand{\R}{\mbox{\hspace{.04mm}\rule{0.2mm}{2.8mm}\hspace{-1.5mm} R}}

\nc{\spa}{\hspace{.1cm},\hspace{1 cm}}
\nc{\vs}{\vspace}
\nc{\NP}[1]{Nucl.\ Phys.\ {\bf #1}}
\nc{\PL}[1]{Phys.\ Lett.\ {\bf #1}}
\nc{\CMP}[1]{Commun.\ Math.\ Phys.\ {\bf #1}}
\nc{\PR}[1]{Phys.\ Rev.\ {\bf #1}}
\nc{\PRL}[1]{Phys.\ Rev.\ Lett.\ {\bf #1}}
\nc{\PTP}[1]{Prog.\ Theor.\ Phys.\ {\bf #1}}
\nc{\PTPS}[1]{Prog.\ Theor.\ Phys.\ Suppl.\ {\bf #1}}
\nc{\MPL}[1]{Mod.\ Phys.\ Lett.\ {\bf #1}}
\nc{\IJMP}[1]{Int.\ Jour.\ Mod.\ Phys.\ {\bf #1}}
\nc{\IM}[1]{Invent.\ Math.\ {\bf #1}}
\nc{\SJNP}[1]{Sov. J. Nucl. Phys.\ {\bf #1}}
\nc{\JHEP}[1]{J.\ High\ Energy Phys.\ {\bf #1}}

\def\vvdots{\mathinner{\mkern1mu\raise1pt\vbox{\kern7pt\hbox{.}}\mkern2mu
 \raise4pt\hbox{.}\mkern2mu\raise7pt\hbox{.}\mkern1mu}}

\begin{document}

\topmargin -1mm
\oddsidemargin 5mm

\begin{titlepage}
\setcounter{page}{0}

\vs{8mm}
\begin{center}
{\huge Higher $su(N)$ tensor products}

\vs{15mm}
{\large J. Rasmussen}\footnote{rasmussj@cs.uleth.ca; supported in part
by a PIMS Postdoctoral Fellowship and by NSERC} and 
{\large M.A. Walton}\footnote{walton@uleth.ca; supported in part by NSERC}
\\[.2cm]
{\em Physics Department, University of Lethbridge,
Lethbridge, Alberta, Canada T1K 3M4}

\end{center}

\vs{8mm}
\centerline{{\bf{Abstract}}}
\noindent
We extend our recent results on ordinary $su(N)$ tensor product multiplicities
to higher $su(N)$ tensor products. Particular emphasis is put on four-point
couplings where the tensor product of four highest weight modules is 
considered. The number of times the singlet occurs in the decomposition is
the associated multiplicity. In this framework, 
ordinary tensor products correspond to three-point couplings.
As in that case, the four-point multiplicity may be expressed
explicitly as a multiple sum measuring the discretised volume of a
convex polytope. This description extends to higher-point couplings
as well. We also address the problem of determining when a higher-point
coupling exists, i.e., when the associated multiplicity is non-vanishing. 
The solution is a set of inequalities in the Dynkin labels.

\end{titlepage}
\newpage
\renewcommand{\thefootnote}{\arabic{footnote}}
\setcounter{footnote}{0}

\section{Introduction}

The decomposition of tensor products of simple Lie algebra modules 
has been studied for a long
time now. Many elegant results have been found for the multiplicities 
of the decompositions, the so-called tensor product multiplicities.  
However, most results pertain to the decomposition of tensor products
of {\em two} irreducible highest weight modules of a simple Lie algebra:
\ben
 M_\la\otimes M_\mu=\bigoplus_\nu\ {T_{\la,\mu}}^\nu M_\nu\ .
\label{MM}
\een
$M_\la$ is the module of highest weight $\la$, while ${T_{\la,\mu}}^\nu$ is
the tensor product multiplicity. This problem is equivalent 
to the more symmetric one of determining the multiplicity of the 
singlet in the decomposition of the {\em triple} product
\ben
 M_\la\otimes M_\mu\otimes M_\nu\supset T_{\la,\mu,\nu}M_0\ .
\label{MMM}
\een
Indeed, if $\nu^+$ denotes the weight conjugate to 
$\nu$, we have $T_{\lambda,\mu,\nu} = {T_{\lambda,\mu}}^{\nu^+}$.  
We will use the shorthand notation $\la\otimes\mu\otimes\nu$ to represent the
left hand side of (\ref{MMM}), and refer to it as a three-point product.

The objective of the present work is to discuss higher $su(N)$ tensor
products (or higher-point $su(N)$ couplings), 
and provide explicit expressions for the associated multiplicities
\ben
 M_\la\otimes M_\mu\otimes...\otimes M_\sigma\supset T_{\la,\mu,...,\sigma}
  M_0\ .
\label{MMMM}
\een

Based on a generalisation of the Berenstein-Zelevinsky method of triangles
\cite{BZ}, we have recently obtained very explicit expressions for
$T_{\la,\mu,\nu}$. The result is a multiple sum formula measuring the 
discretised volume of a convex polytope associated to the tensor product
\cite{RW}. It is this idea which shall be extended here to cover
higher-point couplings. The main focus will be on four-point couplings.
Our results pertain to the $A$-series, $A_r=su(r+1)$.

We also address the problem of determining when a higher-point
coupling exists, i.e., when the associated multiplicity
is non-vanishing. The solution is a set of inequalities in the Dynkin labels.

\section{Ordinary tensor product multiplicities}

To fix notation, we review briefly our main result 
\cite{RW} on the computation of ordinary tensor product multiplicities,
i.e., on the evaluation of three-point couplings. We refer to \cite{RW}
for more details.

An $su(r+1)$ Berenstein-Zelevinsky (BZ) triangle, describing a 
particular coupling (to the singlet) associated to the
triple product $\lambda\otimes\mu\otimes\nu$, is a triangular arrangement of 
\ben
 E_r=\frac{3}{2}r(r+1)
\label{E}
\een
non-negative integers, denoted entries. The entries are subject to
certain constraints: the $3r$ outer constraints and the $2H_r$
hexagon identities, where
\ben
 H_r=\frac{1}{2}r(r-1)
\label{H}
\een
is the number of hexagons, see below.
The case $su(3)$ provides a simple illustration:
\ben
 \matrix{m_{13}\cr
	n_{12}~~\quad l_{23}\cr
 m_{23}~\quad\qquad ~~m_{12}\cr
 n_{13}~\quad l_{12} \qquad n_{23} \quad~ l_{13} \cr }
\label{trithree}
\een
According to the outer constraints, these $E_2=9$ non-negative integers 
are related to the Dynkin labels of the three integrable highest weights by
\ben
\begin{array}{llll}
 &m_{13}+n_{12}=\lambda_1\ ,\ \ &n_{13}+l_{12}=\mu_1\ ,\ \
 &l_{13}+m_{12}=\nu_1\ ,\nn
 &m_{23}+n_{13}=\lambda_2\ ,\ \ &n_{23}+l_{13}=\mu_2\ ,\ \
 &l_{23}+m_{13}=\nu_2\ .
\end{array}
\label{outthree}
\een 
The entries further satisfy the hexagon conditions 
\ben
\begin{array}{l}
 n_{12}+m_{23}=n_{23}+m_{12}\ ,\nn 
 m_{12}+l_{23}=m_{23}+l_{12}\ ,\nn 
 l_{12}+n_{23}=l_{23}+n_{12}\ ;
\end{array}
\label{hexthree}
\een
of which only two are independent. The number of BZ triangles is the
triple tensor product multiplicity $T_{\la,\mu,\nu}$.

The generalisation of the BZ triangles we consider is obtained by
weakening the constraint that all entries are {\it non-negative} integers to
{\it arbitrary} 
integers, negative as well as non-negative. The hexagon identities
and the outer constraints are still enforced. A triangle will be called a
{\it true} BZ triangle if all its entries are non-negative.

A generalised $su(r+1)$ BZ triangle is built out of $H_r$ hexagons and 
three corner points. Each hexagon corresponds to two 
independent constraints on the entries. This leaves
\ben
 E_r-(2H_r+3r)=H_r
\label{EHrel}
\een
parameters labelling the possible triangles. 
Thus, for a given triple product $\la\otimes\mu\otimes\nu$, the set of
associated triangles spans an $H_r$-dimensional lattice.
Among the lattice points, only a finite number correspond to true 
BZ triangles. As already stated, this number 
is precisely the tensor product multiplicity of the triple product.

The $H_r$ basis vectors in the lattice correspond to so-called (basis)
virtual triangles \cite{RW}, denoted ${\cal V}$. 
They are themselves triangles (i.e., points in the lattice) associated to
the particular coupling $0\otimes0\otimes0$. In the case $su(4)$ the three
basis virtual triangles are
\bea
 \matrix{1\cr
	\bar1~~\quad\bar1\cr
 \bar1~\quad\quad ~~\bar1\cr
 1~\quad\bar1\quad\bar1\quad ~1\cr
 0\qquad\quad 1\qquad\quad 0 \cr
 0~\quad 0~\quad 0~\quad 0~\quad 0~\quad 0 \cr}
\hspace{1.5cm} \matrix{0\cr
	0~~\quad 0\cr
 1~\quad\quad ~~0\cr
 \bar1~\quad \bar1\quad 1 \quad ~0\cr
 \bar1\qquad\quad \bar1\qquad\quad 0 \cr
 1~\quad\bar1~\quad\bar1~\quad 1~\quad 0~\quad 0 \cr}
\hspace{1.5cm}
 \matrix{0\cr
	0~~\quad 0\cr
 0~\quad\quad ~~1\cr
 0~\quad 1\quad\bar1 \quad ~\bar1\cr
 0\qquad\quad\bar1\qquad\quad\bar1 \cr
 0~\quad 0~\quad 1~\quad\bar1~\quad\bar1~\quad 1\cr}
\label{Vfour}
\eea
where $\bar 1\equiv -1$.
In general, a convenient basis for the virtual triangles is given by
associating the simple distribution
\ben
 \matrix{\matrix{1\cr 1\quad\bar1~~\quad\bar1\quad 1\cr
  \bar1~\quad\quad ~~\bar1\cr
  1~\quad\bar1~\quad\bar1~\quad 1\cr
 1\cr}}
\label{virt}
\een
of plus and minus ones to any given hexagon. All other entries are zero.

The lattice may now be characterised by an initial triangle ${\cal T}_0$
and the basis of virtual triangles, as a generic triangle ${\cal T}$ may be
written
\ben
 {\cal T}={\cal T}_0+\sum_{i,j\geq1}^{i+j=r}v_{i,j}{\cal V}_{i,j}\ .
\label{T}
\een
$v_{i,j}$ are called linear coefficients, and our choice of labelling
follows from\footnote{The choice of labelling differs slightly from the
one used in Ref. \cite{RW}.}
\ben
\matrix{\star\cr
	\star~~\qquad \star\cr
 \star\qquad v_{r-1,1}\quad \star\cr
 \star\qquad\ \star\qquad\ \ \star\ \qquad \star\cr
 \star~\quad\ v_{r-2,1}\quad \star\quad\ v_{r-2,2}\quad~\star \cr
 \star\quad\ \ \star\qquad \ \star\quad~~\quad \star\qquad\ \star\quad\ \ 
  \star \cr
 \vvdots\ 
  \qquad\qquad\qquad\vdots\ \ \ \qquad\vdots\qquad\qquad\qquad\ddots\cr
 \star \quad\qquad\qquad\qquad\qquad\quad
   \qquad\qquad\qquad\qquad\qquad \star\cr
	\star~~\qquad \star \ \quad\qquad\qquad\qquad\qquad\quad
   \qquad\qquad\qquad \star~~\qquad \star\cr
 \star~\qquad v_{2,1}\quad ~~\star
   \qquad\qquad\qquad \dots \qquad\qquad\qquad
 \star~\quad v_{2,r-2}\quad ~~\star\cr
 \star~\quad\ \ \star~~\qquad \star \ \qquad \star 
    \qquad\qquad\qquad\qquad\qquad
 \star\qquad\ \star~~\qquad \star\ \ \quad ~\star\cr
 \star\qquad v_{1,1} \ \quad \star\qquad v_{1,2}\ \ \quad \star 
     \qquad\quad\qquad\qquad\
 \star\quad\ \ v_{1,r-2}\quad \star\quad\ \ v_{1,r-1}\quad \star \cr
 \star\qquad \star\ \ \quad \star~~\qquad \star\qquad \ \star\qquad~ 
  \star  
     \quad\ \dots \quad\
 \star\qquad~ \star\qquad \star~~\qquad \star\qquad \star\qquad 
  \star \cr\cr}
\label{hexnot}
\een
The initial triangle corresponds to any point in the lattice.
A convenient choice is based on the fact that every highest weight 
$\nu$ in a coupling $\lambda\otimes\mu\otimes\nu$ satisfies 
\ben
 \nu\ =\ \lambda^++\mu^+ -\sum_{i=1}^rn_i\alpha_i\ ,\ \ \ \ \
  n_i  \ =\ (\lambda^+)^i +(\mu^+)^i - \nu^i\ \in\ \Z_\ge\ ,
\label{Nal}
\een
where $\alpha_i$ is the $i$-th simple root. The coefficients $n_i$ are 
expressed using dual Dynkin labels: 
\ben
 \la\ =\ \sum_{i=1}^r\la_i\Lambda^i\ =\ \sum_{i=1}^r\la^i\al_i^\vee\ , 
\label{Dynkin}
\een
where $\{\Lambda^i\}$ and $\{\al^\vee_i\}$ are the sets of fundamental
weights and simple co-roots, respectively. For simply-laced algebras
like $su(N)$, $\alpha_i$ is identical to the co-root $\alpha_i^\vee$ (with  
standard normalisation $\alpha^2 = 2$, for $\alpha$ a long root). 
Now, it is easy to construct the unique true triangle associated to the 
coupling $\la\otimes\mu\otimes(\la+\mu)^+$, as well as triangles associated to
the couplings $0\otimes0\otimes\al_i$. $su(3)$ examples are 
\ben
 \la\otimes\mu\otimes(\la+\mu)^+\ \ \sim\ \  
 \matrix{\lambda_1\cr
	0~~\quad \mu_1\cr
 \lambda_2~\quad\qquad ~~\lambda_2\cr
 0~\quad \mu_1 \qquad 0 \quad~ \mu_2 \cr }
 \hspace{1.5cm}
  0\otimes0\otimes\alpha_2\ \ \sim\ \ \matrix{1\cr
	\bar 1~~\quad 1\cr
 0~\quad\qquad ~~\bar1\cr
 0~\quad 0 \qquad 0 \quad~ 0 \cr}
\label{alii}
\een
Adding the triangles according to (\ref{Nal}) results in a BZ triangle
associated to $\lambda\otimes\mu\otimes\nu$.
The result for $su(r+1)$ is the following generalised BZ triangle:
\ben
 \matrix{ N'_{r} \cr
      n_r \qquad N_{r} \cr
 \ \ \la_{2} \qquad\qquad\ \ \ \ N'_{r-1}  \cr
 \ \ 0 \ \qquad \mu_{1} \qquad n_{r-1} \quad N_{r-1} \cr
 \ \la_{3} \quad\qquad\qquad\ \ \la_{3} \quad\qquad\qquad N'_{r-2} \cr
 \ \ 0\ \qquad\ \mu_{1}\qquad 0\qquad\ \ \ \mu_2\quad\ \ n_{r-2}
   \ \ \ N_{r-2} \cr
 \ \vvdots \quad\qquad\qquad\qquad \vdots\qquad\quad \ \
  \vdots\qquad\qquad\qquad\quad \ddots\ \ \  \cr
 \la_{r-2}  \qquad\qquad\qquad\qquad\qquad\quad
   \qquad\qquad\qquad\qquad\qquad  N'_{3} \ \      \cr
 0\ \qquad\ \mu_{1}  \qquad\qquad\qquad\qquad\qquad\qquad\qquad\qquad
   \qquad\quad \     n_{3}\qquad N_{3} \ \   \cr
 \la_{r-1}\qquad\qquad\ \la_{r-1} \qquad\qquad\qquad\ \dots\qquad 
  \qquad\qquad\quad 
    \la_{r-1}\ \qquad\qquad\ N'_{2}\quad  \cr
 0\qquad\ \mu_{1}\qquad 0\qquad\ \mu_{2}
   \qquad\qquad\qquad\qquad\qquad \quad\qquad\
  0\qquad \mu_{r-2}\qquad n_{2}\ \ \ \quad N_{2} \    \cr  
 \ \la_{r}\qquad\qquad\quad\ \la_{r}\qquad\qquad\quad \la_{r}
    \qquad\qquad\qquad\qquad\quad \ \ 
  \la_{r}\qquad\qquad\quad\ \la_{r}\qquad\quad\quad\ \ \ N'_{1}\quad \cr  
 0\qquad\ \mu_{1}\qquad 0\quad~\quad \mu_2\quad~~~ 0\qquad\ \ \mu_{3} 
   \qquad\quad\dots\qquad\quad
 0\qquad\mu_{r-2}\ \ \ \ 0\quad\ \ \
  \mu_{r-1}\quad\ \ n_{1}\quad\ \ N_{1}\ \ \cr\cr}
\label{initial}
\een
The entries $n_i$, $N_i$ and $N'_i$ are defined by
\bea
 n_i&=&\la^{r-i+1}+\mu^{r-i+1}-\nu^i\ ,\nn
 N_i&=&(1-\delta_{i1})n_{i-1}-n_i+\mu_{r-i+1}\nn
  &=&-\la^{r-i+1}+(1-\delta_{i1})
  \la^{r-i+2}-(1-\delta_{ir})\mu^{r-i}
  +\mu^{r-i+1}-(1-\delta_{i1})\nu^{i-1}+\nu^i\ ,\nn
 N'_i&=&\nu_i-N_i\nn
  &=&\la^{r-i+1}-(1-\delta_{i1})\la^{r-i+2}+(1-\delta_{ir})\mu^{r-i}
  -\mu^{r-i+1}+\nu^i-(1-\delta_{ir})\nu^{i+1}\ .
\label{nN}
\eea
supplemented by the condition that $n_i$, and thus also $N_i$ and $N_i'$, 
are integers (cf. (\ref{Nal})).

A true BZ triangle associated to the product $\la\otimes\mu\otimes\nu$
has the additional property that all its entries
are {\em non-negative} integers. From (\ref{T}) it then follows that
the set of true BZ triangles may be characterised by a set of inequalities 
in the linear coefficients. It is easily seen that these
inequalities define a convex
polytope embedded in $\R^{H_r}$, whose discretised volume is the
tensor product multiplicity $T_{\la,\mu,\nu}$. As discussed in Ref. \cite{RW},
this volume may be measured explicitly and calculated by a multiple sum:
\ben
 T_{\la,\mu,\nu}=\left(\sum_{v_{1,1}}\right)\left(\sum_{v_{2,1}}\sum_{v_{1,2}}
   \right)...\left(\sum_{v_{r-2,1}}...\sum_{v_{1,r-2}}\right)
    \left(\sum_{v_{r-1,1}}...\sum_{v_{1,r-1}}\right)1\ .
\label{vol}
\een
Here the summation variables are bounded according to
\bea
 &&{\rm max}\{-N_1,v_{1,r-2},-N_2'+v_{2,r-2},-\mu_{r-2}
  +v_{1,r-2}-v_{2,r-3}+v_{2,r-2}\}\nn
 &&\quad\quad
  \leq v_{1,r-1}\leq{\rm min}\{n_1,\mu_{r-1}+v_{1,r-2},\la_r-v_{1,r-2}
  +v_{2,r-2},\nn
 &&\hspace{3.3cm}
  n_2+v_{1,r-2}-v_{2,r-2},N_1',N_2+v_{2,r-2}\}\ ,\nn
 &&{\rm max}\{v_{l-1,r-l}-v_{l-1,r-l-1}+v_{l,r-l-1},-N_{l+1}'
   +v_{l+1,r-l-1},\nn
 &&\quad\quad\quad-\mu_{r-l-1}+v_{l+1,r-l-1}
   -(1-\delta_{l,r-2})v_{l+1,r-l-2}+v_{l,r-l-1}\}\nn
 &&\quad\quad
  \leq v_{l,r-l}\leq{\rm min}\{\la_{r-l+1}-v_{l,r-l-1}
   +v_{l-1,r-l}+v_{l+1,r-l-1},\nn
 &&\hspace{3.2cm}n_{l+1}+v_{l,r-l-1}-v_{l+1,r-l-1},N_{l+1}+v_{l+1,r-l-1}\}\ ,\ 
  \ {\rm for}\ 2\leq l\leq r-2\ ,\nn
 &&{\rm max}\{v_{r-2,1},-N_r'\}
 \leq v_{r-1,1}\leq{\rm min}\{\la_2+v_{r-2,1},n_r,N_r\}\ ,\nn
 &&{\rm max}\{v_{1,j-1},-\mu_{j-1}+v_{1,j-1}+v_{2,j-1}-(1-\delta_{j,2})
  v_{2,j-2}\}\nn
 &&\quad\quad
   \leq v_{1,j}\leq{\rm min}\{\mu_{j}+v_{1,j-1},
  \la_r-v_{1,j-1}+v_{2,j-1}\}\ ,\ \ {\rm for}\ 2\leq j\leq r-2\ ,\nn
 &&{\rm max}\{v_{i,j-1}+v_{i-1,j}-v_{i-1,j-1},-\mu_{j-1}
  +v_{i,j-1}+v_{i+1,j-1}-(1-\delta_{j,2})v_{i+1,j-2}\}\nn
 &&\quad\quad\leq v_{i,j}\leq\la_{r-i+1}-v_{i,j-1}+v_{i+1,j-1}+v_{i-1,j}
  \ ,\ \ {\rm for}\ 2\leq i,j,i+j\leq r-1\ ,\nn
 &&v_{i-1,1}\leq v_{i,1}\leq
  \la_{r-i+1}+v_{i-1,1}\ ,\ \ {\rm for}\ 2\leq i\leq r-2\ ,\nn
 &&0\leq v_{1,1}\leq{\rm min}\{\mu_1,\la_r\}\ .
\label{bounds}
\eea
Due to the freedom in choosing the initial triangle, 
this is just one out of an infinite class of multiple sum representations
of $T_{\la,\mu,\nu}$. Its asymmetry in the weights merely reflects the 
symmetry-breaking choice of initial triangle, and the choice of an
order of summation. 

\section{Four-point products}

We focus first on four-point products:
\ben
 M_\la\otimes M_\mu\otimes M_\nu\otimes M_\sigma\supset T_{\la,\mu,\nu,\sigma}
  M_0\ .
\label{Mfour}
\een
The objective is to characterise the multiplicity $T_{\la,\mu,\nu,\sigma}$
as the discretised volume of a convex polytope, and then to express
the volume explicitly as a multiple sum. Our starting point will be the
decomposition
\ben
 T_{\la,\mu,\nu,\sigma}=\sum_\rho T_{\la,\mu,\rho}T_{\rho^+,\nu,\sigma}
\label{TTT}
\een
which may be represented graphically by
\ben
\mbox{
\begin{picture}(100,80)(50,-35)
\unitlength=1cm
\thicklines
\put(-1,1){\line(1,-1){2}}
\put(-1,-1){\line(1,1){2}}
\put(2.4,0){=}
\put(3.4,0){$\sum_\rho$}
\put(5,1){\line(1,-1){1}}
\put(5,-1){\line(1,1){1}}
\put(6,0){\line(1,0){1.5}}
\put(7,0){\line(-2,1){0.4}}
\put(7,0){\line(-2,-1){0.4}}
\put(7.5,0){\line(1,1){1}}
\put(7.5,0){\line(1,-1){1}}
\put(-1.3,1.2){$\la$}
\put(-1.3,-1.3){$\mu$}
\put(1.2,-1.3){$\nu$}
\put(1.2,1.2){$\sigma$}
\put(4.7,1.2){$\la$}
\put(4.7,-1.3){$\mu$}
\put(8.7,-1.3){$\nu$}
\put(8.7,1.2){$\sigma$}
\put(6.7,0.4){$\rho$}
\end{picture}
}
\label{rho}
\een
The arrow indicates that the third weight associated to the coupling
involving $\la$ and $\mu$ is $\rho$, while its conjugate, $\rho^+$,
takes part in the second coupling.
Due to the $S_4$ symmetry of $T_{\la,\mu,\nu,\sigma}$ there are many possible
decompositions in terms of three-point couplings, associated to 
different channels. As in the case of the three-point couplings, we are not
seeking a symmetric representation of the multiplicity, but merely an
explicit multiple sum formula. 
The decomposition (\ref{TTT}) is itself a representation but less
explicit than our goal. The former corresponds to considering a sum over 
products of discretised volumes of convex polytopes embedded in
$H_r$-dimensional euclidean spaces. The sum is over the $r$ Dynkin labels
of the interior weight $\rho$. Our goal is to represent the multiplicity
as the discretised volume of a {\em single} convex polytope embedded in a
$(2H_r+r)$-dimensional euclidean space, and eventually to measure explicitly
the volume in terms of a multiple sum. The number of parameters is
of course conserved.

Our approach does not admit an immediate and simple modification in order
to obtain a result which is manifestly $S_4$ symmetric.
Taking an average over the possible (symmetry-breaking)
channels would provide a straightforward symmetrisation but result
in a much more complicated expression.
One should rather look for an approach which respects the four-coupling
nature, and thus does not rely on breaking the coupling up into three-point
couplings. 

Let 
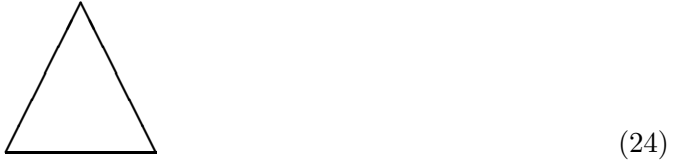
\ben
\mbox{
\begin{picture}(100,70)(-25,0)
\unitlength=1cm
\thicklines
 \put(0,0){\line(1,2){1}}
 \put(0,0){\line(1,0){2}}
 \put(2,0){\line(-1,2){1}}
\end{picture}
}
\een
denote a BZ triangle whose entries have not been specified explicitly.
This offers an alternative illustration of the channel (\ref{rho}):
\ben
\mbox
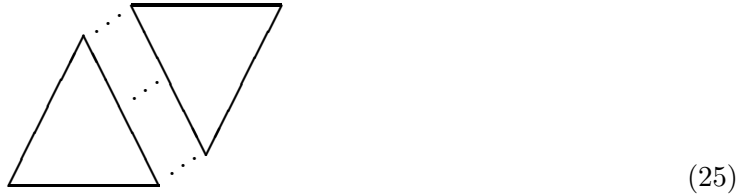
{
\begin{picture}(100,80)(0,0)
\unitlength=1cm
\thicklines
 \put(0,0){\line(1,2){1}}
 \put(0,0){\line(1,0){2}}
 \put(2,0){\line(-1,2){1}}
 \put(1.1,2){$\vvdots$}
 \put(2.1,0.1){$\vvdots$}
 \put(1.6,1.1){$\vvdots$}
 \put(1.5,2.4){
\begin{picture}(50,50)
 \put(0,0){\line(1,-2){1}}
 \put(0,0){\line(1,0){2}}
 \put(2,0){\line(-1,-2){1}}
\end{picture}}
\end{picture}
}
\label{four}
\een
The dotted lines indicate a gluing of the triangles. These composite objects
may be regarded as a generalisation of the BZ triangles to
diagrams associated to particular choices of four-point channels; 
governed by a gluing of triangles representing three-point couplings.
Disregarding this origin of (\ref{four}), the configuration is merely
an arrangement of $2E_r$ (non-)negative integers subject to certain
constraints: $4H_r$ hexagon identities, $4r$ outer constraints
representing the four weights, and $r$ gluing constraints.
Along the dotted lines, the original $2r$ outer constraints are
substituted by the $r$ gluing constraints requiring opposite weights to be
identical. A four-point diagram is called true if all entries are
{\em non-negative} integers. 
Explicit examples of four-point diagrams are provided below.

The number of parameters labelling the possible four-point diagrams is
\ben
 2E_r-(4H_r+4r+r)=2H_r+r\ .
\label{fourpara}
\een
As in the three-point case, this reflects the existence of $2H_r+r$
basis virtual diagrams that correspond to the basis vectors in the 
$(2H_r+r)$-dimensional lattice associated to any given four-point product
$\la\otimes\mu\otimes\nu\otimes\sigma$. The points in the lattice are 
the four-point diagrams. 

A triangle consisting of zeros alone is called a zero-triangle.
It is then obvious that $2H_r$ of the basis virtual diagrams are made up of
a basis virtual triangle glued together with a zero-triangle, while
the remaining $r$ virtual diagrams are associated to the gluing.
We shall denote virtual diagrams of the first kind by ${\cal V}^{(1)}_{i,j}$ 
or ${\cal V}^{(2)}_{i,j}$ (depending on which triangle includes the non-trivial
part)\footnote{The lower indices $i$ and $j$ refer to any chosen labelling
of the $H_r$ hexagons and, thus, of the associated virtual triangles.
(\ref{hexnot}) is a merely a convenient choice.} and call them extended (basis)
virtual triangles, and virtual diagrams of the second kind by 
${\cal G}_i$ and call them ``simple gluing roots''.
It follows from (\ref{Nal}) that the latter indeed correspond to pairs
of simple roots, as illustrated by this $su(3)$ example (cf. (\ref{alii})):
\ben
\mbox{
\begin{picture}(100,100)(125,0)
\unitlength=1cm
\thicklines
\put(-1,1.5){${\cal G}_1\ \ =$}
\put(0,0){0}
\put(1,0){0}
\put(2,0){$\bar1$}
\put(3,0){1}
\put(0.5,0.7){0}
\put(2.5,0.7){1}
\put(1,1.4){0}
\put(2,1.4){$\bar1$}
\put(1.5,2.1){0}
\put(1.8,2.4){$\vvdots$}
\put(2.3,1.7){$\vvdots$}
\put(2.8,1){$\vvdots$}
\put(3.3,0.3){$\vvdots$}
\put(2.4,2.7){0}
\put(3.4,2.7){0}
\put(4.4,2.7){0}
\put(5.4,2.7){0}
\put(2.9,2.0){$\bar1$}
\put(4.9,2){0}
\put(3.4,1.3){1}
\put(4.4,1.3){$\bar1$}
\put(3.9,0.6){1}
\put(7.2,0){\begin{picture}(100,100)
\put(-1,1.5){${\cal G}_2\ \ =$}
\put(0,0){0}
\put(1,0){0}
\put(2,0){0}
\put(3,0){0}
\put(0.5,0.7){0}
\put(2.5,0.7){$\bar1$}
\put(1,1.4){$\bar1$}
\put(2,1.4){1}
\put(1.5,2.1){1}
\put(1.8,2.4){$\vvdots$}
\put(2.3,1.7){$\vvdots$}
\put(2.8,1){$\vvdots$}
\put(3.3,0.3){$\vvdots$}
\put(2.4,2.7){1}
\put(3.4,2.7){$\bar1$}
\put(4.4,2.7){0}
\put(5.4,2.7){0}
\put(2.9,2.0){1}
\put(4.9,2){0}
\put(3.4,1.3){$\bar1$}
\put(4.4,1.3){0}
\put(3.9,0.6){0}
\end{picture}}
\end{picture}
}
\label{GG}
\een

A natural generalisation of conventional notation allows us to represent
graphically the gluing roots as tree-graphs:
\ben
\mbox{
\begin{picture}(100,100)(230,-40)
\unitlength=1cm
\thicklines
\put(3,0){${\cal G}_1\ \ \sim$}
\put(5,1){\line(1,-1){1}}
\put(5,-1){\line(1,1){1}}
\put(6,0){\line(1,0){1.5}}
\put(7,0){\line(-2,1){0.4}}
\put(7,0){\line(-2,-1){0.4}}
\put(7.5,0){\line(1,1){1}}
\put(7.5,0){\line(1,-1){1}}
\put(4.7,1.2){0}
\put(4.7,-1.3){0}
\put(8.7,-1.3){0}
\put(8.7,1.2){0}
\put(6.6,0.4){$\al_1$}
\put(7.2,0){\begin{picture}(100,100)
\put(3,0){${\cal G}_2\ \ \sim$}
\put(5,1){\line(1,-1){1}}
\put(5,-1){\line(1,1){1}}
\put(6,0){\line(1,0){1.5}}
\put(7,0){\line(-2,1){0.4}}
\put(7,0){\line(-2,-1){0.4}}
\put(7.5,0){\line(1,1){1}}
\put(7.5,0){\line(1,-1){1}}
\put(4.7,1.2){0}
\put(4.7,-1.3){0}
\put(8.7,-1.3){0}
\put(8.7,1.2){0}
\put(6.6,0.4){$\al_2$}
\end{picture}}
\end{picture}
}
\een
These graphs, of course, represent couplings that do not exist (i.e., they
have vanishing multiplicities) but nevertheless serve to illustrate
the power of virtual diagrams. One may even extend the summation range
in (\ref{rho}) to include them, since
the associated algebraic expressions (\ref{TTT}) merely contribute
zeros whenever non-true diagrams are encountered.

As in the three-point case, we may now characterise any diagram 
${\cal D}$ in the lattice by specifying an initial diagram ${\cal D}_0$:
\ben
 {\cal D}={\cal D}_0+\sum_{a=1,2}\sum_{i,j\geq1}^{i+j=r}v_{i,j}^{(a)}
 {\cal V}_{i,j}^{(a)}-\sum_{i=1}^rg_i{\cal G}_i\ .
\label{DVG}
\een
$v_{i,j}^{(a)}$ and $g_i$ are integers, denoted linear coefficients.
Note that we have chosen a convention with a minus sign in front
of the last sum (\ref{DVG}). This reflects the intimate relationship between 
a simple gluing root and (our convention (\ref{virt}) for)
a basis virtual triangle:
\ben
\mbox{
\begin{picture}(100,120)(70,-50)
\unitlength=1cm
\thicklines
\put(-2.8,0){$-{\cal V}\ \ \ =$}
\put(0,0){$1$}
\put(2.4,0){$1$}
\put(-0.6,0.9){$\bar1$}
\put(0.6,0.9){$1$}
\put(1.8,0.9){$1$}
\put(3,0.9){$\bar1$}
\put(1.2,1.8){$\bar1$}
\put(-0.6,-0.9){$\bar1$}
\put(0.6,-0.9){$1$}
\put(1.8,-0.9){$1$}
\put(3,-0.9){$\bar1$}
\put(1.2,-1.8){$\bar1$}
 \put(8,0){\begin{picture}(50,50)
\put(-2.8,0){${\cal G}\ \ \ =$}
\put(0,0){$1$}
\put(2.4,0){$1$}
\put(-0.6,0.9){$\bar1$}
 \put(0.8,0.5){$\vvdots$}
\put(1.8,0.9){$1$}
\put(3,0.9){$\bar1$}
\put(1.2,1.8){$\bar1$}
\put(-0.6,-0.9){$\bar1$}
\put(0.6,-0.9){$1$}
 \put(1.5,-0.6){$\vvdots$}
\put(3,-0.9){$\bar1$}
\put(1.2,-1.8){$\bar1$}
\end{picture}}
\end{picture}
}
\label{VG}
\een
One may of course choose to substitute the gluing roots with virtual
triangles. However, that would introduce redundant parameters that cannot 
be fixed. This is because such extended gluing roots have the same number of
entries as virtual triangles but fewer constraints.
$su(2)$ offers a simple illustration of that where the single gluing root
is substituted by a single hexagon:
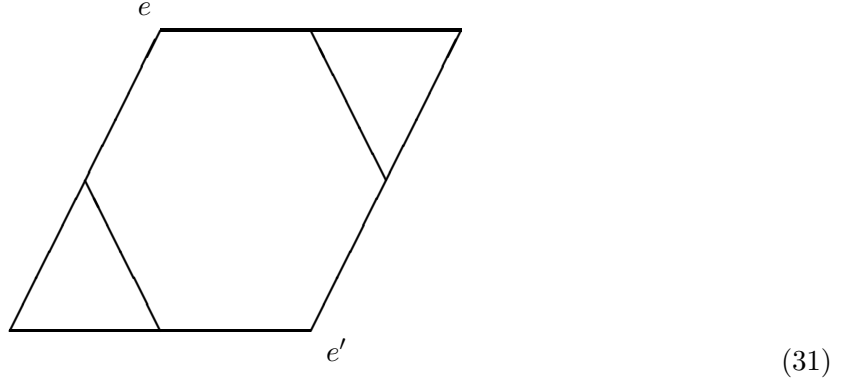
\ben
\mbox{
\begin{picture}(100,150)(35,-15)
\unitlength=1cm
\thicklines
 \put(0,0){\line(1,2){2}}
\put(4,0){\line(1,2){2}}
 \put(0,0){\line(1,0){4}}
\put(2,4){\line(1,0){4}}
 \put(2,0){\line(-1,2){1}}
\put(5,2){\line(-1,2){1}}
\put(4.2,-0.4){$e'$}
\put(1.7,4.2){$e$}
\end{picture}
}
\label{red}
\een
The redundancy occurs since the hexagon identities allow us to fix
only a relation between the two extra entries $e$ and $e'$.
This ``uniformisation'' makes possible relations between
higher-point $su(N')$ couplings and
lower-point $su(N)$ couplings, when $N$ is sufficiently larger than
$N'$. The four-point $su(2)$ diagram above may thus be embedded
in a three-point $su(N\geq4)$ diagram. We hope to discuss such relations in
the future. Here, however, we refrain from replacing the gluing roots. 

Now we turn to the construction of a convenient initial diagram ${\cal D}_0$.
Referring to (\ref{TTT}) we know that
\bea
 \rho&=&\la^++\mu^+-\sum_{i=1}^rn_i^{(1)}\al_i\ ,\ \ \ \ \ n_i^{(1)}=
  (\la^+)^i+(\mu^+)^i-\rho^i\ \in\ \Z_\geq\ ,\nn
 \rho^+&=&\nu^++\sigma^+-\sum_{i=1}^rn_i^{(2)}\al_i\ ,\ \ \ \ \ n_i^{(2)}=
  (\nu^+)^i+(\sigma^+)^i-(\rho^+)^i\ \in\ \Z_\geq\ .
\label{rhorho}
\eea
It follows that 
\ben
 \la+\mu+\nu+\sigma\ =\ \sum_{i=1}^rm_i\al_i\ ,\ \ m_i\ \in\ \Z_\geq\ ,
\label{selfour}
\een
ensuring the integer nature of the entries.
This invites us to choose
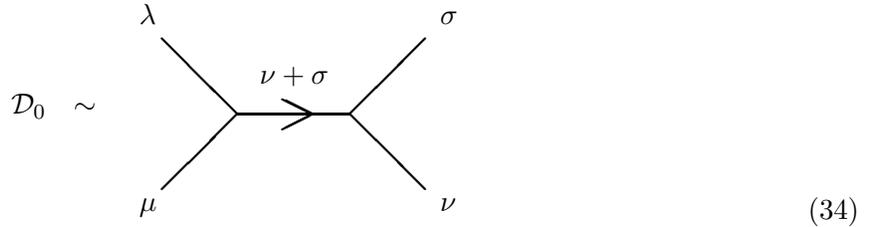
\ben
\mbox{
\begin{picture}(100,100)(130,-40)
\unitlength=1cm
\thicklines
\put(3,0){${\cal D}_0\ \ \sim$}
\put(5,1){\line(1,-1){1}}
\put(5,-1){\line(1,1){1}}
\put(6,0){\line(1,0){1.5}}
\put(7,0){\line(-2,1){0.4}}
\put(7,0){\line(-2,-1){0.4}}
\put(7.5,0){\line(1,1){1}}
\put(7.5,0){\line(1,-1){1}}
\put(4.7,1.2){$\la$}
\put(4.7,-1.3){$\mu$}
\put(8.7,-1.3){$\nu$}
\put(8.7,1.2){$\sigma$}
\put(6.3,0.4){$\nu+\sigma$}
\end{picture}
}
\label{initialD}
\een
as the initial diagram. This ${\cal D}_0$ is easily constructed explicitly
as it corresponds to gluing together our original initial triangles
${\cal T}_0$ (\ref{initial})
associated to the couplings $\la\otimes\mu\otimes(\nu+\sigma)$
and $\nu\otimes\sigma\otimes(\nu+\sigma)^+$, respectively.

Now, requiring that ${\cal D}$ in (\ref{DVG}) is a true diagram leads
to a set of inequalities in the linear coefficients $v$ and $g$. It follows
from the structure of the virtual diagrams that this set defines a
convex polytope in the euclidean space $\R^{2H_r+r}=\R^{r^2}$.
The discretised volume of the polytope is by construction the tensor
product multiplicity $T_{\la,\mu,\nu,\sigma}$. This characterisation 
of the four-point tensor product multiplicity is our first main result.

To measure the volume in a straightforward manner, we should organise
the inequalities such that a multiple sum expressing the volume
may be written down without having to evaluate intersections of polytope
faces. This corresponds to choosing an
``appropriate'' order of summation, as discussed in Ref. \cite{RW}.
Anticipating the extension to higher tensor products to be discussed below,
we propose the following procedure.

Let the left (or lower) triangle in (\ref{four}) correspond initially
to the product $\la\otimes\mu\otimes(\nu+\sigma)$ such that the right
triangle initially corresponds to the product
$\nu\otimes\sigma\otimes(\nu+\sigma)^+$. These couplings are of course
altered by the gluing process, adding linear combinations of roots to
the third weights. Denoting the linear coefficients of the virtual
triangles $v_{i,j}^{(1)}$ and $v_{i,j}^{(2)}$, respectively, we may
choose the labelling indicated in the following diagram (cf. (\ref{hexnot}))
\ben
\mbox
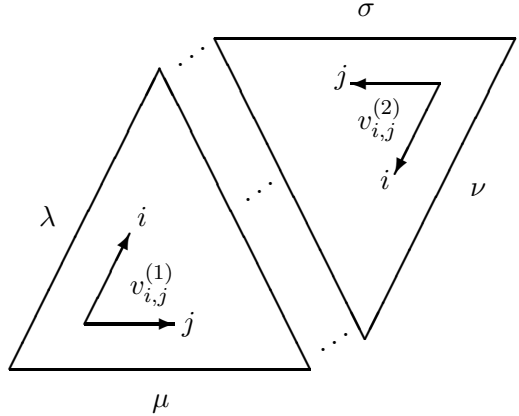
{
\begin{picture}(100,190)(50,-25)
\unitlength=1cm
\thicklines
 \put(0,0){\line(1,2){2}}
 \put(0,0){\line(1,0){4}}
 \put(4,0){\line(-1,2){2}}
 \put(2.2,4){$\vvdots$}
 \put(4.1,0.2){$\vvdots$}
 \put(3.1,2.2){$\vvdots$}
 \put(0.4,1.9){$\la$}
 \put(1.9,-0.5){$\mu$}
 \put(1,0.6){\vector(1,2){0.6}}
 \put(1,0.6){\vector(1,0){1.2}}
 \put(1.6,1.0){$v_{i,j}^{(1)}$}
 \put(1.7,1.9){$i$}
 \put(2.3,0.5){$j$}
 \put(2.6,4.4){
\begin{picture}(50,50)
 \put(0,0){\line(1,-2){2}}
 \put(0,0){\line(1,0){4}}
 \put(4,0){\line(-1,-2){2}}
 \put(1.9,0.3){$\sigma$}
 \put(3.4,-2.1){$\nu$}
 \put(3,-0.6){\vector(-1,0){1.2}}
 \put(3,-0.6){\vector(-1,-2){0.6}}
 \put(2.2,-2){$i$}
 \put(1.6,-0.6){$j$}
 \put(1.9,-1.2){$v_{i,j}^{(2)}$}
\end{picture}}
\end{picture}
}
\label{fourorder}
\een
An appropriate order of summation is then obtained by starting with
the right-most variable, $v_{1,1}^{(2)}$, and moving systematically
towards left:
\bea
 T_{\la,\mu,\nu,\sigma}&=&\left(\sum_{v_{1,1}^{(1)}}\right)
  \left(\sum_{v_{2,1}^{(1)}}\sum_{v_{1,2}^{(1)}}\right)...
  \left(\sum_{v_{r-1,1}^{(1)}}...\sum_{v_{1,r-1}^{(1)}}\right)
  \left(\sum_{g_r}...\sum_{g_1}\right)\nn
 &\times&\left(\sum_{v_{1,r-1}^{(2)}}...\sum_{v_{r-1,1}^{(2)}}\right)
  ...\left(\sum_{v_{1,2}^{(2)}}\sum_{v_{2,1}^{(2)}}\right)
  \left(\sum_{v_{1,1}^{(2)}}\right)1\ .
\label{sumfour}
\eea
The summation variables are bounded according to
\bea
 &&{\rm max}\{0,-\sigma_2+v^{(2)}_{1,2},v^{(2)}_{1,2}-v^{(2)}_{2,2}+
  v^{(2)}_{2,1},-\nu_{r-1}+v^{(2)}_{2,1}\}\nn
 &&\quad\quad
   \leq v^{(2)}_{1,1}\leq{\rm min}\{\sigma_1,v^{(2)}_{1,2},\nu_r-
  v^{(2)}_{1,2}+v^{(2)}_{2,1},\sigma_1+v^{(2)}_{1,2}-v^{(2)}_{2,1},
  v^{(2)}_{2,1},\nu_r\}\ ,\nn
 &&{\rm max}\{-\sigma_2+v^{(2)}_{i-1,2}-v^{(2)}_{i-1,3}+v^{(2)}_{i,2},\nn
   &&\hspace{2cm}
   v^{(2)}_{i,2}-(1-\delta_{i,r-2})v^{(2)}_{i+1,2}-\delta_{i,r-2}g_2
  +v^{(2)}_{i+1,1},-\nu_{i+1}+v^{(2)}_{i+1,1}\}\nn
 &&\quad\quad
   \leq v^{(2)}_{i,1}\leq{\rm min}\{
   \nu_{r-i+1}+v^{(2)}_{i-1,2}-v^{(2)}_{i,2}+v^{(2)}_{i+1,1},\nn
   &&\hspace{3cm}\sigma_1+v^{(2)}_{i,2}-v^{(2)}_{i+1,1},v^{(2)}_{i+1,1}\}
        \ ,\ \ {\rm for}\ 2\leq i\leq r-2\ ,\nn
 &&{\rm max}\{v^{(2)}_{i+1,j}+v^{(2)}_{i,j+1}-(1-\delta_{i+j,r-1})
   v^{(2)}_{i+1,j+1}-\delta_{i+j,r-1}g_{j+1},\nn
  &&\hspace{2cm}-\sigma_{j+1}+
   v^{(2)}_{i-1,j+1}-v^{(2)}_{i-1,j+2}+v^{(2)}_{i,j+1}\}\nn
 &&\quad\quad
   \leq v^{(2)}_{i,j}\leq\nu_{r-i+1}+v^{(2)}_{i-1,j+1}-v^{(2)}_{i,j+1}
   +v^{(2)}_{i+1,j}\ ,\ \ {\rm for}\ 2\leq i,j,i+j\leq r-1\ ,\nn
 &&{\rm max}\{-\sigma_{j+1}+v^{(2)}_{1,j+1},v^{(2)}_{1,j+1}+v^{(2)}_{2,j}
  -(1-\delta_{j,r-2})v^{(2)}_{2,j+1}-\delta_{j,r-2}g_{r-1}\}\nn
 &&\quad\quad
   \leq v^{(2)}_{1,j}\leq{\rm min}\{v^{(2)}_{1,j+1},\nu_r-v^{(2)}_{1,j+1}
    +v^{(2)}_{2,j}\}\ ,\ \ {\rm for}\ 2\leq j\leq r-2\ ,\nn
 &&{\rm max}\{-\nu_1+g_1,-\sigma_2+v^{(2)}_{r-2,2}+g_2-g_3\}\nn
 &&\quad\quad
   \leq v^{(2)}_{r-1,1}\leq{\rm min}\{\sigma_1-g_1+g_2,\nu_2+v^{(2)}_{r-2,2}
    +g_1-g_2,g_1\}\ ,\nn
 &&-\sigma_{l+1}+v^{(2)}_{r-l-1,l+1}+g_{l+1}-g_{l+2}\nn
 &&\quad\quad
   \leq v^{(2)}_{r-l,l}\leq\nu_{l+1}-v^{(2)}_{r-l-1,l+1}+g_l-g_{l+1}
    \ ,\ \ {\rm for}\ 2\leq l\leq r-2\ ,\nn
 &&-\sigma_r+g_r\leq v^{(2)}_{1,r-1}\leq{\rm min}\{g_r,\nu_r+g_{r-1}-g_r\}
  \ ,\nn
 &&{\rm max}\{-n_1+v^{(1)}_{1,r-1},-N_2+v^{(1)}_{1,r-1}-v^{(1)}_{2,r-2}
    +g_2\}\nn
 &&\quad\quad
   \leq g_1\leq{\rm min}\{N_1+v^{(1)}_{1,r-1},N_1'-v^{(1)}_{1,r-1}+g_2\}
  \ ,\nn
 &&{\rm max}\{-n_i+v^{(1)}_{i-1,r-i+1}+v^{(1)}_{i,r-i}-v^{(1)}_{i-1,r-i},
  -N_{i+1}+v^{(1)}_{i,r-i}-(1-\delta_{i,r-1})v^{(1)}_{i+1,r-i-1}+g_{i+1}\}\nn
 &&\quad\quad
   \leq g_i\leq N_i'+v^{(1)}_{i-1,r-i+1}-v^{(1)}_{i,r-i}+g_{i+1}\}
  \ ,\ \ {\rm for}\ 2\leq i\leq r-1\ ,\nn
 &&-n_r+v^{(1)}_{r-1,1}\leq g_r\leq N_r'+v^{(1)}_{r-1,1}\ ,\nn
 &&{\rm max}\{v^{(1)}_{1,j-1},-\mu_{j-1}+v^{(1)}_{1,j-1}
  +v^{(1)}_{2,j-1}-(1-\delta_{j,2})v^{(1)}_{2,j-2}\}\nn
 &&\quad\quad
   \leq v^{(1)}_{1,j}\leq{\rm min}\{\mu_{j}+v^{(1)}_{1,j-1},
  \la_r-v^{(1)}_{1,j-1}+v^{(1)}_{2,j-1}\}
   \ ,\ \ {\rm for}\ 2\leq j\leq r-1\ ,\nn
 &&{\rm max}\{v^{(1)}_{i,j-1}+v^{(1)}_{i-1,j}-v^{(1)}_{i-1,j-1},-\mu_{j-1}
  +v^{(1)}_{i,j-1}+v^{(1)}_{i+1,j-1}-(1-\delta_{j,2})v^{(1)}_{i+1,j-2}\}\nn
 &&\quad\quad\leq v^{(1)}_{i,j}\leq\la_{r-i+1}-v^{(1)}_{i,j-1}
  +v^{(1)}_{i+1,j-1}+v^{(1)}_{i-1,j}\ ,\ \ {\rm for}\ 2\leq i,j,i+j\leq r\ ,\nn
 &&v^{(1)}_{i-1,1}\leq v^{(1)}_{i,1}\leq
  \la_{r-i+1}+v^{(1)}_{i-1,1}\ ,\ \ {\rm for}\ 2\leq i\leq r-1\ ,\nn
 &&0\leq v^{(1)}_{1,1}\leq{\rm min}\{\mu_1,\la_r\}\ ,
\label{fourbound}
\eea
where the parameters $n_i$, $N_i$ and $N_i'$ are defined as in 
(\ref{nN}), with $\nu$ replaced by $\nu+\sigma$:
\bea
 n_i&=&\la^{r-i+1}+\mu^{r-i+1}-(\nu+\sigma)^i\ ,\nn
 N_i&=&(1-\delta_{i1})n_{i-1}-n_i+\mu_{r-i+1}\nn
  &=&-\la^{r-i+1}+(1-\delta_{i1})
  \la^{r-i+2}-(1-\delta_{ir})\mu^{r-i}
  +\mu^{r-i+1}-(1-\delta_{i1})(\nu+\sigma)^{i-1}+(\nu+\sigma)^i\ ,\nn
 N'_i&=&(\nu+\sigma)_i-N_i\nn
  &=&\la^{r-i+1}-(1-\delta_{i1})\la^{r-i+2}+(1-\delta_{ir})\mu^{r-i}
  -\mu^{r-i+1}+(\nu+\sigma)^i-(1-\delta_{ir})(\nu+\sigma)^{i+1}
\label{nNfour}
\eea
This multiple sum formula is our second main result.
For $su(2)$, $su(3)$ and $su(4)$ the explicit multiple sum formulas are
provided in Section 5.

\section{Higher point couplings}

We shall now indicate how one may extend our results
on four-point couplings to any higher ${\cal N}$-point coupling,
in straightforward fashion.

It is well-known that higher-point couplings may be decomposed into 
three-point couplings along the lines of (\ref{TTT}).
The various tree-graph channels all have diagram counterparts as 
illustrated by the following two examples:
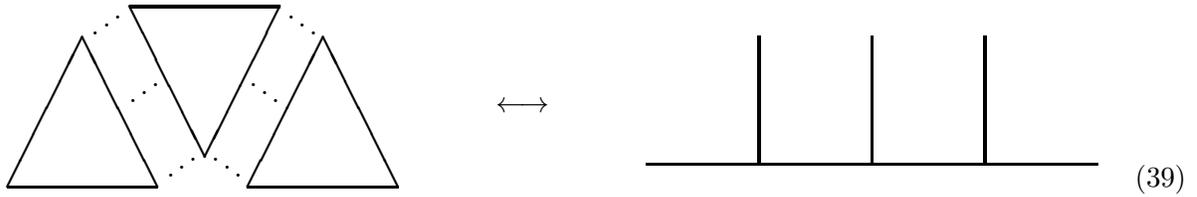
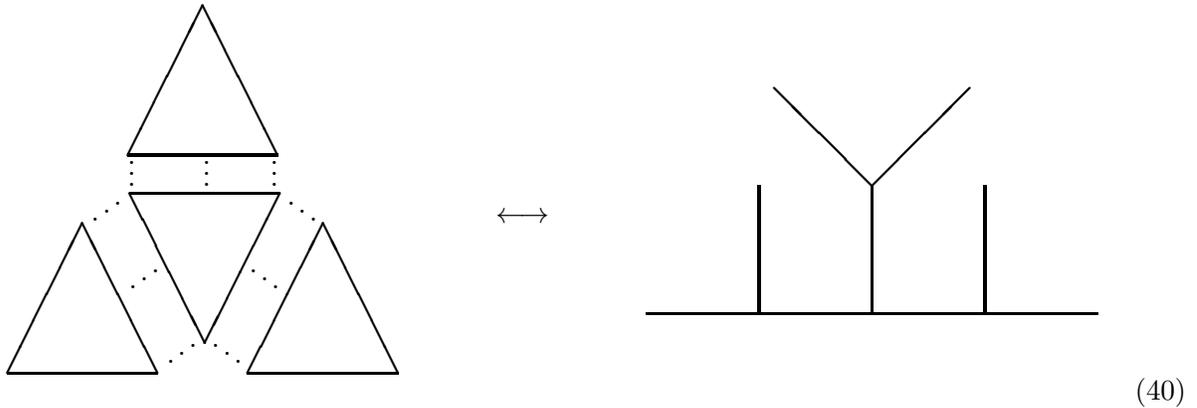
\ben
\mbox{
\begin{picture}(100,100)(170,0)
\unitlength=1cm
\thicklines
 \put(0,0){\line(1,2){1}}
 \put(0,0){\line(1,0){2}}
 \put(2,0){\line(-1,2){1}}
 \put(1.1,2){$\vvdots$}
 \put(2.1,0.1){$\vvdots$}
 \put(1.6,1.1){$\vvdots$}
 \put(1.5,2.4){
\begin{picture}(50,50)
 \put(0,0){\line(1,-2){1}}
 \put(0,0){\line(1,0){2}}
 \put(2,0){\line(-1,-2){1}}
\end{picture}}
\put(2.7,0.1){$\ddots$}
\put(3.7,2){$\ddots$}
\put(3.2,1.1){$\ddots$}
\put(3.2,0){\begin{picture}(50,50)
 \put(0,0){\line(1,2){1}}
 \put(0,0){\line(1,0){2}}
 \put(2,0){\line(-1,2){1}}
\end{picture}}
\put(6.5,1){$\longleftrightarrow$}
\put(8.5,0.3){\begin{picture}(50,50)
 \put(0,0){\line(1,0){6}}
 \put(1.5,0){\line(0,1){1.7}}
 \put(3,0){\line(0,1){1.7}}
 \put(4.5,0){\line(0,1){1.7}}
\end{picture}}
\end{picture}
}
\label{string}
\een
and
\ben
\mbox{
\begin{picture}(100,170)(170,-10)
\unitlength=1cm
\thicklines
 \put(0,0){\line(1,2){1}}
 \put(0,0){\line(1,0){2}}
 \put(2,0){\line(-1,2){1}}
 \put(1.1,2){$\vvdots$}
 \put(2.1,0.1){$\vvdots$}
 \put(1.6,1.1){$\vvdots$}
 \put(1.5,2.4){
\begin{picture}(50,50)
 \put(0,0){\line(1,-2){1}}
 \put(0,0){\line(1,0){2}}
 \put(2,0){\line(-1,-2){1}}
\end{picture}}
\put(2.7,0.1){$\ddots$}
\put(3.7,2){$\ddots$}
\put(3.2,1.1){$\ddots$}
\put(3.2,0){\begin{picture}(50,50)
 \put(0,0){\line(1,2){1}}
 \put(0,0){\line(1,0){2}}
 \put(2,0){\line(-1,2){1}}
\end{picture}}
\put(1.6,2.5){$\vdots$}
\put(2.6,2.5){$\vdots$}
\put(3.5,2.5){$\vdots$}
\put(1.6,2.9){\begin{picture}(50,50)
 \put(0,0){\line(1,2){1}}
 \put(0,0){\line(1,0){2}}
 \put(2,0){\line(-1,2){1}}
\end{picture}}
\put(6.5,2){$\longleftrightarrow$}
\put(8.5,0.8){\begin{picture}(50,50)
 \put(0,0){\line(1,0){6}}
 \put(1.5,0){\line(0,1){1.7}}
 \put(3,0){\line(0,1){1.7}}
 \put(4.5,0){\line(0,1){1.7}}
 \put(3,1.7){\line(1,1){1.3}}
 \put(3,1.7){\line(-1,1){1.3}}
\end{picture}}
\end{picture}
}
\label{rocket}
\een
Since we may choose the channel freely, we can avoid complicated 
configurations like the ``rocket'' of 
(\ref{rocket}) and concentrate on the ``string-like'' ones, 
like (\ref{string}) and the following nine-point coupling:
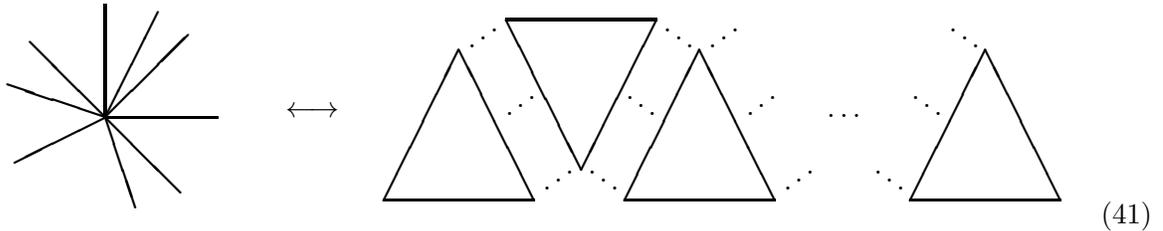
\ben
\mbox{
\begin{picture}(100,100)(120,-40)
\unitlength=1cm
\thicklines
\put(0,0){\line(1,0){1.5}}
\put(0,0){\line(1,1){1.1}}
\put(0,0){\line(1,2){0.7}}
\put(0,0){\line(0,1){1.5}}
\put(0,0){\line(-1,1){1}}
\put(0,0){\line(-3,1){1.3}}
\put(0,0){\line(-2,-1){1.2}}
\put(0,0){\line(1,-3){0.4}}
\put(0,0){\line(1,-1){1}}
\put(2.4,0){$\longleftrightarrow$}
\put(3.7,-1.1){\begin{picture}(50,50)
 \put(0,0){\line(1,2){1}}
 \put(0,0){\line(1,0){2}}
 \put(2,0){\line(-1,2){1}}
 \put(1.1,2){$\vvdots$}
 \put(2.1,0.1){$\vvdots$}
 \put(1.6,1.1){$\vvdots$}
 \put(1.5,2.4){
\begin{picture}(50,50)
 \put(0,0){\line(1,-2){1}}
 \put(0,0){\line(1,0){2}}
 \put(2,0){\line(-1,-2){1}}
\end{picture}}
\put(2.7,0.1){$\ddots$}
\put(3.7,2){$\ddots$}
\put(3.2,1.1){$\ddots$}
\put(3.2,0){\begin{picture}(50,50)
 \put(0,0){\line(1,2){1}}
 \put(0,0){\line(1,0){2}}
 \put(2,0){\line(-1,2){1}}
\end{picture}}
 \put(4.3,2){$\vvdots$}
 \put(5.3,0.1){$\vvdots$}
 \put(4.8,1.1){$\vvdots$}
\put(5.9,1.1){$\dots$}
\put(7,0){\begin{picture}(50,50)
 \put(0,0){\line(1,2){1}}
 \put(0,0){\line(1,0){2}}
 \put(2,0){\line(-1,2){1}}
\end{picture}}
\put(6.5,0.1){$\ddots$}
\put(7.5,2){$\ddots$}
\put(7,1.1){$\ddots$}
\end{picture}}
\end{picture}
}
\label{string2}
\een
Thus, an ${\cal N}$-point coupling may conveniently be represented
by an ${\cal N}$-point diagram consisting of ${\cal N}-2$
triangles glued together along ${\cal N}-3$ pairs of faces to form a
string-like configuration. An ${\cal N}$-point diagram 
is therefore a geometrical arrangement of $({\cal N}-2)E_r$ 
(non-)negative integers subject to $2({\cal N}-2)H_r$ hexagon
identities, ${\cal N}r$ outer constraints, and $({\cal N}-3)r$
gluing constraints. (This is also true for the more complicated diagrams,
such as (\ref{rocket})). This leaves
\ben
 ({\cal N}-2)E_r-\left(({\cal N}-2)H_r+{\cal N}r+({\cal N}-3)r\right)
  =({\cal N}-2)H_r+({\cal N}-3)r
\label{para}
\een
parameters labelling the possible diagrams.
As it should be, this is equal to the total number of virtual triangles and
simple gluing roots.

Thus, from the point of view of the ${\cal N}$-point diagram, we have
two types of virtual diagrams: extended (basis) virtual triangles
${\cal V}$, and simple gluing roots ${\cal G}$, exacly as for
four-point couplings.
The extension of (\ref{DVG}) is therefore obvious:
\ben
 {\cal D}={\cal D}_0+\sum_{a=1}^{{\cal N}-2}
  \sum_{i,j\geq1}^{i+j=r}v_{i,j}^{(a)}
  {\cal V}_{i,j}^{(a)}-\sum_{a=1}^{{\cal N}-3}\sum_{i=1}^rg_i^{(a)}
  {\cal G}_i^{(a)}\ .
\label{DVG2}
\een
The initial diagram ${\cal D}_0$ is likewise easy to describe, as
it may be constructed by gluing ${\cal N}-2$ initial triangles together.
Labelling the ${\cal N}$ weights according to (in this example 
${\cal N}$ is assumed odd)
\ben
\mbox{
\begin{picture}(100,120)(90,-20)
\unitlength=1cm
\thicklines
 \put(0,0){\line(1,2){1}}
 \put(0,0){\line(1,0){2}}
 \put(2,0){\line(-1,2){1}}
\put(-0.2,1){$\la^{(1)}$}
\put(0.8,-0.5){$\la^{(2)}$}
 \put(1.1,2){$\vvdots$}
 \put(2.1,0.1){$\vvdots$}
 \put(1.6,1.1){$\vvdots$}
 \put(1.5,2.4){
\begin{picture}(50,50)
 \put(0,0){\line(1,-2){1}}
 \put(0,0){\line(1,0){2}}
 \put(2,0){\line(-1,-2){1}}
\put(0.7,0.2){$\la^{({\cal N})}$}
\end{picture}}
\put(2.7,0.1){$\ddots$}
\put(3.7,2){$\ddots$}
\put(3.2,1.1){$\ddots$}
\put(3.2,0){\begin{picture}(50,50)
 \put(0,0){\line(1,2){1}}
 \put(0,0){\line(1,0){2}}
 \put(2,0){\line(-1,2){1}}
\put(0.8,-0.5){$\la^{(3)}$}
\end{picture}}
 \put(4.3,2){$\vvdots$}
 \put(5.3,0.1){$\vvdots$}
 \put(4.8,1.1){$\vvdots$}
\put(5.9,1.1){$\dots$}
\put(7,0){\begin{picture}(50,50)
 \put(0,0){\line(1,2){1}}
 \put(0,0){\line(1,0){2}}
 \put(2,0){\line(-1,2){1}}
\put(0.2,-0.5){$\la^{(({\cal N}+1)/2)}$}
\put(1.7,1){$\la^{(({\cal N}+3)/2)}$}
\end{picture}}
\put(6.5,0.1){$\ddots$}
\put(7.5,2){$\ddots$}
\put(7,1.1){$\ddots$}
\end{picture}
}
\label{lll}
\een
the participating initial triangles are associated to the couplings
\bea
 &&\la^{(({\cal N}+1)/2)}\otimes\la^{(({\cal N}+3)/2)}\otimes
  (\la^{(({\cal N}+1)/2)}+\la^{(({\cal N}+3)/2)})^+\ ,\nn
 &&(\la^{(({\cal N}+1)/2)}+\la^{(({\cal N}+3)/2)})\otimes
  \la^{(({\cal N}+5)/2)}\otimes
  (\la^{(({\cal N}+1)/2)}+\la^{(({\cal N}+3)/2)}+\la^{(({\cal N}+5)/2)})^+
   \ ,\nn
 &&\vdots\nn
 &&(\la^{(3)}+...+\la^{({\cal N}-1)})\otimes\la^{({\cal N})}\otimes
  (\la^{(3)}+...+\la^{({\cal N})})^+\ ,\nn
 &&\la^{(1)}\otimes\la^{(2)}\otimes(\la^{(3)}+...+\la^{({\cal N})})\ .
\label{ts}
\eea
The weights are subject to the consistency condition (cf. (\ref{selfour}))
\ben 
 \la^{(1)}+...+\la^{({\cal N})}=\sum_{i=1}^rm_i\al_i\ ,\ \ \ m_i\ \in\ \Z_\geq
   \ .
\label{selN}
\een

The characterisation of the associated tensor product multiplicity
in terms of a convex polytope, is materialised by requiring
that the diagram should be a true diagram, i.e., all entries must
be {\em non-negative} integers. As before, its discretised volume
is the multiplicity by construction. That volume can be expressed explicitly
as a multiple sum. An appropriate order of summation is indicated here:
\ben
 T_{\la^{(1)},\la^{(2)},...,\la^{({\cal N})}}
 =\{\sum_{v^{(1)}}\}\{\sum_{g^{(1)}}\}...
  \{\sum_{v^{({\cal N}-3)}}\}\{\sum_{g^{({\cal N}-3)}}\}
 \{\sum_{v^{({\cal N}-2)}}\}1\ .
\label{sumN}
\een
This generalisation of our main results on three- and four-point couplings,
concludes the extension to general higher-point couplings.

\section{Examples and an application}

It is of interest to know whether or not an ${\cal N}$-point coupling 
$\la\otimes\mu\otimes...
\otimes\sigma$ exists, without having to work out the tensor product 
multiplicity. Based on our multiple sum formulas 
(\ref{sumfour}) and (\ref{sumN}), 
one may derive a set of inequalities in the dual and ordinary Dynkin
labels of the ${\cal N}$ weights, determining when the associated tensor 
product multiplicity is non-vanishing. The method
is an immediate extension of the one employed in Ref.
\cite{RW} when discussing three-point couplings (\ref{vol}). We work out the
inequalities for $su(2)$ and $su(3)$ four-point couplings.
In principle, it is possible to repeat the procedure for higher rank
and higher ${\cal N}$ than four, though it rapidly becomes
cumbersome.

To the best of our knowledge, similar results only exist for three-point
products where, besides our work \cite{RW}, the works
\cite{Zel,Ful} provide recent results and extensive lists of references.

For $su(2)$ the BZ triangle representing the product
$\la\otimes\mu\otimes\nu$ is unique ($H_1=0$):
\ben
\mbox{
\begin{picture}(80,80)(50,-15)
\unitlength=1cm
\thicklines
\put(0,0){$\hf(\la_1+\mu_1-\nu_1)$}
\put(3.6,0){$\hf(-\la_1+\mu_1+\nu_1)$}
\put(1.8,1.5){$\hf(\la_1-\mu_1+\nu_1)$}
\end{picture}
}
\een
Nevertheless, gluing two triangles together leaves one free parameter $g$. 
We have
\ben
 {\cal D}={\cal D}_0-g{\cal G}
\label{DDG}
\een
where
\ben
\mbox{
\begin{picture}(80,130)(110,-15)
\unitlength=1cm
\thicklines
\put(-2,1.5){${\cal D}_0\ \ =$}
\put(0,0){$\hf(\la_1+\mu_1-\nu_1-\sigma_1)$}
\put(4.5,0){$\hf(-\la_1+\mu_1+\nu_1+\sigma_1)$}
\put(2.2,1.5){$\hf(\la_1-\mu_1+\nu_1+\sigma_1)$}
\put(9,0.6){$\vvdots$}
\put(7.2,2.1){$\vvdots$}
\put(10,1.5){$\nu_1$}
\put(8.5,3){$\sigma_1$}
\put(11.5,3){$0$}
\end{picture}
}
\een
and
\ben
\mbox{
\begin{picture}(100,100)(-45,-40)
\unitlength=1cm
\thicklines
\put(-2.8,0){${\cal G}\ \ \ =$}
\put(0,0){$1$}
\put(2.4,0){$1$}
 \put(0.8,0.5){$\vvdots$}
\put(1.8,0.9){$1$}
\put(3,0.9){$\bar1$}
\put(-0.6,-0.9){$\bar1$}
\put(0.6,-0.9){$1$}
 \put(1.5,-0.6){$\vvdots$}
\end{picture}
}
\een

Requiring ${\cal D}$ to be a true diagram results in a set of inequalities
defining a one-dimensional convex polytope - a line segment.
Its discretised volume (or length) is the sought multiplicity:
\ben
 T_{\la,\mu,\nu,\sigma}=\sum_{g={\rm max}\{0,\ S-\la_1-\mu_1\}}^{{\rm min}\{
  S-\la_1,\ S-\mu_1,\ \nu_1,\ \sigma_1\}}1
   \ ,\ \ \ \ \ S\equiv\hf(\la_1+\mu_1+\nu_1
     +\sigma_1)\ \in\ \Z_\geq\ .
\label{Tsu2}
\een
The summation, and thus the multiplicity, is non-vanishing if and only if
the upper bound is greater than or equal to the lower bound.
This requirement defines a four-dimensional cone:
\ben
 0\leq\la_1,\ \mu_1,\ \nu_1,\ \sigma_1,\ S-\la_1,\ S-\mu_1,\ S-\nu_1,\ 
   S-\sigma_1\ .
\label{conesu2}
\een
It is easily verified that (\ref{Tsu2}) and (\ref{conesu2}) are in
accordance with well-known results. 

For $su(3)$ the four-point coupling may be characterised by a convex
polytope in a four-dimensional euclidean space. Its discretised volume
is the tensor product multiplicity which we find may be expressed as
the following multiple sum:
\bea
 T_{\la,\mu,\nu,\sigma}&=&\sum_{v^{(1)}=0}^{{\rm min}\{\la_2,\ \mu_1\}}\ 
 \sum_{g_2=-n_2+v^{(1)}}^{N_2'+v^{(1)}}\ 
 \sum_{g_1={\rm max}\{-N_2+v^{(1)}+g_2,\ -n_1+v^{(1)}\}}^{
  {\rm min}\{N_1+v^{(1)},\ N_1'+g_2-v^{(1)}\}}\nn
 &\times&
 \sum_{v^{(2)}={\rm max}\{0,\ -\sigma_2+g_2,\ -\nu_1+g_1\}}^{
  {\rm min}\{\nu_2,\ \sigma_1,\ g_1,\ g_2,\ \nu_2+g_1-g_2,\ \sigma_1
    -g_1+g_2\}}1\ ,
\eea
where the weights are subject to 
 \ben
 S_i\equiv\la^i+\mu^i+\nu^i+\sigma^i\ \in\ \Z_\geq\ ,\ \ \ \ i=1,2
\een
and where
\bea
 n_1&=&\la^2+\mu^2-\nu^1-\sigma^1\ ,\nn
 n_2&=&\la^1+\mu^1-\nu^2-\sigma^2\ ,\nn
 N_1&=&-\la^2-\mu^1+\mu^2+\nu^1+\sigma^1\ ,\nn
 N_2&=&-\la^1+\la^2+\mu^1-\nu^1+\nu^2-\sigma^1+\sigma^2\ ,\nn
 N_1'&=&\la^2+\mu^1-\mu^2+\nu^1-\nu^2+\sigma^1-\sigma^2\ ,\nn
 N_2'&=&\la^1-\la^2-\mu^1+\nu^2+\sigma^2\ .
\eea
This explicit result is believed to be new.

Analysing when the tensor product multiplicity is non-vanishing, leads to
the following definition of a cone in the eight-dimensional Dynkin label
space:
\bea
 0&\leq&\la_i,\ \mu_i,\ \nu_i,\ \sigma_i\ ,\ \ \ \ i=1,2\nn
 0&\leq&S_i-\la_1-\la_2,\ S_i-\mu_1-\mu_2,\ S_i-\nu_1-\nu_2,\ 
    S_i-\sigma_1-\sigma_2
  \ ,\ \ \ \ i=1,2\nn
 0&\leq&S_i-\la_i-\mu_i,\ S_i-\la_i-\nu_i,\ S_i-\la_i-\sigma_i,\nn
 && S_i-\mu_i-\nu_i,\ S_i-\mu_i-\sigma_i,\ S_i-\nu_i-\sigma_i\ ,\ \ \ \ i=1,2
\label{cone3}
\eea
This explicit characterisation is also believed to be new.
It is verified immediately that for one weight equal to zero,
(\ref{cone3}) reduces to the result
for the three-point product discussed in Ref. \cite{RW}, i.e.,
$T_{\la,\mu,\nu,0}>0$ if and only if $T_{\la,\mu,\nu}>0$.

For ease of use of the formula (\ref{sumfour}) expressing
the tensor product multiplicity $T_{\la,\mu,\nu,\sigma}$ as a multiple sum,
we conclude this section by writing down explicitly the result for $su(4)$: 
\bea
 T_{\la,\mu,\nu,\sigma}&=&\sum_{v^{(1)}_{1,1}=0}^{{\rm min}\{\la_3,\ \mu_1\}}\ 
  \sum_{v^{(1)}_{2,1}=v^{(1)}_{1,1}}^{\la_2+v^{(1)}_{1,1}}\ 
  \sum_{v^{(1)}_{1,2}={\rm max}\{-\mu_1+v^{(1)}_{2,1}+v^{(1)}_{1,1},\ 
  v^{(1)}_{1,1}\}}^{{\rm min}\{\la_3+v^{(1)}_{2,1}-v^{(1)}_{1,1},\ 
   \mu_2+v^{(1)}_{1,1}\}}\ 
  \sum_{g_3=-n_3+v^{(1)}_{2,1}}^{N_3'+v^{(1)}_{2,1}}\nn  
 &\times& \sum_{g_2={\rm max}\{-N_3+g_3+v^{(1)}_{2,1},\ -n_2+v^{(1)}_{1,2}
  +v^{(1)}_{2,1}-v^{(1)}_{1,1}\}}^{N_2'+g_3+v^{(1)}_{1,2}-v^{(1)}_{2,1}}\  
  \sum_{g_1={\rm max}\{-n_1+v^{(1)}_{1,2},\ -N_2+g_2+v^{(1)}_{1,2}
   -v^{(1)}_{2,1}\}}^{{\rm min}\{N_1'+g_2-v^{(1)}_{1,2},\ N_1
    +v^{(1)}_{1,2}\}}\nn 
 &\times&\sum_{v^{(2)}_{1,2}=-\sigma_3+g_3}^{{\rm min}\{\nu_3+g_2-g_3,
   \ g_3\}}\  
  \sum_{v^{(2)}_{2,1}={\rm max}\{-\nu_1+g_1,\ -\sigma_2+v^{(2)}_{1,2}
    +g_2-g_3\}}^{{\rm min}\{\nu_2+v^{(2)}_{1,2}+g_1-g_2,\ \sigma_1-g_1
    +g_2,\ g_1\}}\nn 
 &\times&\sum_{v^{(2)}_{1,1}={\rm max}\{0,\ -\nu_2+v^{(2)}_{2,1}, \
   -\sigma_2+v^{(2)}_{1,2}
   ,\ v^{(2)}_{2,1}+v^{(2)}_{1,2}-g_2\}}^{{\rm min}\{\nu_3,\ \sigma_1,\ 
   v^{(2)}_{2,1},\ v^{(2)}_{1,2},\ \nu_3+v^{(2)}_{2,1}-v^{(2)}_{1,2},\ 
  \sigma_1-v^{(2)}_{2,1}+v^{(2)}_{1,2}\}}\ 1\ .
\eea
The parameters $n_i$, $N_i$ and $N_i'$ are defined in (\ref{nNfour}), while 
the weights are subject to the condition
\ben
 \la^i+\mu^i+\nu^i+\sigma^i\ \in\ \Z_\geq\ ,\ \ \ \ i=1,2,3\ .
\een

\section{Conclusion}

We have generalised our recent work on three-point products \cite{RW}
to cover general ${\cal N}$-point products. That is, we have 
characterised the associated higher tensor product multiplicities
by certain convex polytopes, and measured explicitly their
discretised volumes. The latter are the multiplicities
and are expressed as multiple sums.

The characterisation of the multiplicity as the number of integer points 
in a convex polytope is an example of a polyhedral combinatorial expression.
Alternative polyhedral combinatorial expressions for three-point products
(including other simple Lie algebras as well) may be found in 
\cite{GZ,BZmath}. To the best of our knowledge, our result for higher-point
products is the first of its kind. 

As an application we have also addressed the problem of determining
when a tensor product multiplicity is non-vanishing, and
as an illustration of the general 
resolution provided explicit characterisations for $su(2)$ and $su(3)$.
The result for $su(3)$ is believed to be new.

We are currently extending our work (presented here and in Ref. \cite{RW})
on tensor product multiplicities to fusion multiplicities.
The latter are relevant to the representation theory of affine extensions
of the Lie algebra, the so-called affine Kac-Moody algebras.
They have found prominent applications to conformal field theory
with affine Lie group symmetry, the so-called WZW theories.
Since tensor product multiplicities correspond to the infinite-level limit
of fusion multiplicities, our current efforts are concentrated on 
incorporating the finite-level dependence into the characterisation
of the multiplicities in terms of convex polytopes and their 
discretised volumes. That again relies on our recent studies of
three-point correlation functions in WZW theory \cite{Ras,RW2}.
We intend to report more on this in the future.

Related in spirit to our approach is the recent work Ref. \cite{FW} 
on fusion rules in $SU(N)$ WZW theory. For lower ranks the authors discuss
a combinatorial relation between three-point 
fusion multiplicities and numbers of certain group theoretical orbits.
It would be interesting to understand how the results of Ref. \cite{FW}
are related to ours. 
\\[.3cm]
{\it Acknowledgements}
\\[.2cm]
We thank J. Patera for discussions and T. Gannon for comments.

\end{document}